# Autonomous Cyber Resilience via a Co-Evolutionary Arms Race within a Fortified Digital Twin Sandbox

Malikussaid
*School of Computing, Telkom University*
Bandung, Indonesia
malikussaid@student.telkomuniversity.ac.id

Sutiyo
*School of Computing, Telkom University*
Bandung, Indonesia
tioatmadja@telkomuniversity.ac.id

***Abstract*—The convergence of Information Technology and Operational Technology has exposed Industrial Control Systems to adaptive, intelligent adversaries that render static defenses obsolete. This paper introduces the Adversarial Resilience Co-evolution (ARC) framework, addressing the "Trinity of Trust" comprising model fidelity, data integrity, and analytical resilience. ARC establishes a co-evolutionary arms race within a Fortified Secure Digital Twin (F-SCDT), where a Deep Reinforcement Learning "Red Agent" autonomously discovers attack paths while an ensemble-based "Blue Agent" is continuously hardened against these threats. Experimental validation on the Tennessee Eastman Process (TEP) and Secure Water Treatment (SWaT) testbeds demonstrates superior performance in detecting novel attacks, with $F_1$-scores improving from 0.65 to 0.89 and detection latency reduced from over 1200 seconds to 210 seconds. A comprehensive ablation study reveals that the co-evolutionary process itself contributes a 27% performance improvement. By integrating Explainable AI and proposing a Federated ARC architecture, this work presents a necessary paradigm shift toward dynamic, self-improving security for critical infrastructure.**

## I. The Imperative for Dynamic Resilience in Critical Infrastructure

The convergence of Information Technology (IT) and Operational Technology (OT) has dismantled the "air gap" that once protected critical infrastructure, creating a hyper-connected ecosystem with an expanded attack surface. This vulnerability stems from a philosophical conflict between IT's rapid development model and OT's extreme risk aversion. The introduction of AI has ignited a dynamic arms race requiring a shift from static security models toward resilient, adaptive paradigms.

We propose a "Trinity of Trust" as the foundation for robust Industrial Control System (ICS) security, where failure in any pillar renders the entire security posture invalid:

1. Model Fidelity: The digital representation must accurately reflect the physical asset with quantifiable precision.
2. Data Integrity: The data stream connecting physical to digital must be verifiably authentic and tamper-proof.
3. Analytical Resilience: The anomaly detection models must withstand intelligent adversaries who actively seek to evade them.

This Trinity directly maps to our framework components: F-SCDT addresses Model Fidelity and Data Integrity (Section III), while ARC addresses Analytical Resilience (Section IV), with experimental validation demonstrating measurable improvements in all three pillars (Section V).

### A. *The Foundational Flaws of the OT Environment*

Legacy protocols like Modbus [1] and DNP3 [2] lack basic security features: no authentication, no encryption, and no integrity checks. This creates a "trust-by-default" environment where valid commands are accepted without verification. The 2000 Maroochy Shire sewage spill [3] demonstrated these vulnerabilities, but was treated as an outlier rather than a systemic problem.

### B. *Threat Evolution and Seminal ICS Attacks*

Modern threats range from sophisticated state actors to opportunistic criminals. Groups like Volt Typhoon [4] and APT44 target critical infrastructure, establishing footholds for potential disruption. Significant attacks include:

1. Stuxnet (2010) [5]: Targeted Iranian uranium enrichment centrifuges, manipulating rotational frequencies while replaying normal data to operators.
2. Ukrainian Power Grid (2015) [6]: A coordinated assault that locked operators out while attackers remotely opened breakers, causing widespread outages.
3. Cyber Av3ngers Campaign (2023) [7]: Targeted water systems by exploiting basic security lapses, defacing Human-Machine Interface (HMI) to erode operator trust.

These attacks show a progression from technical exploitation to psychological warfare against operators, requiring defenses that incorporate human factors [8].

### C. *Research Contributions: The ARC Framework*

This paper makes the following contributions:

1. Proposes the ARC framework depicted in Fig. 1 for autonomous, closed-loop hardening of ICS defenses that directly addresses the Trinity of Trust.
2. Formalizes discovery of stealthy, physically-plausible attacks within a Deep Reinforcement Learning (DRL) paradigm with novel reward functions.

3. Provides experimental validation on multiple testbeds with comprehensive ablation studies demonstrating quantifiable improvements in all three trust pillars.
4. Introduces a vision for scaling via Federated Learning (F-ARC) with Byzantine-resilient aggregation.

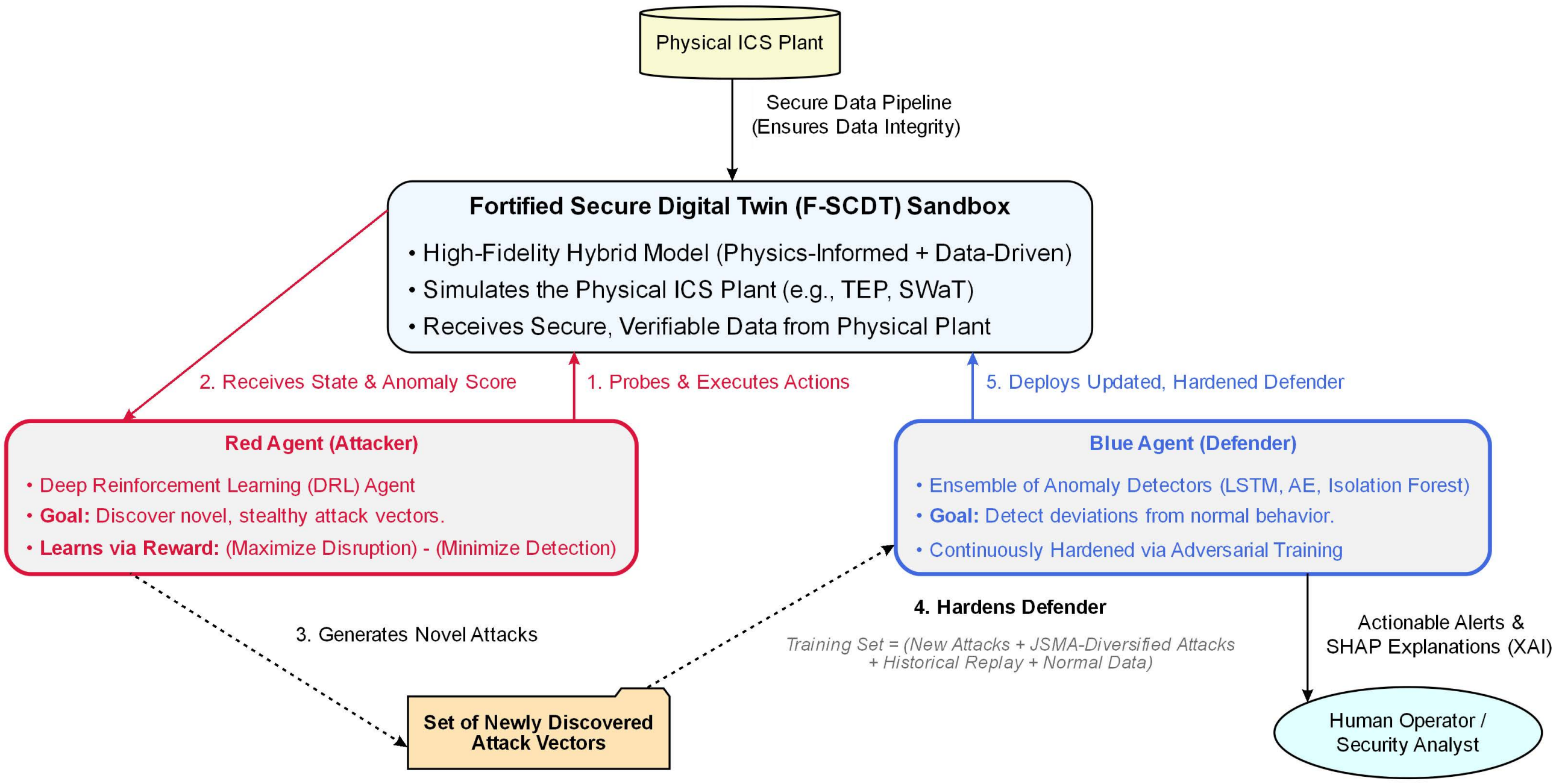

Fig. 1. High-Level Architecture of the ARC Co-Evolutionary Loop

## II. State-of-the-Art in ICS Defense

### A. The Failure of Static Defenses

Early ICS security solutions imported IT concepts like firewalls and signature-based Intrusion Detection Systems (IDS) [9]. These are fundamentally reactive and cannot understand the context of legitimate commands used maliciously. Modern adversaries like Volt Typhoon leverage "living-off-the-land" techniques [10], using legitimate tools to avoid triggering signatures.

Compared to recent Byzantine-resilient approaches [11], static defenses lack the adaptive capability necessary for modern threat landscapes. While blockchain-based integrity solutions [12] address data provenance, they cannot detect semantically valid but malicious commands.

### B. Machine Learning for Anomaly Detection

Machine learning approaches [13] model normal behavior to detect deviations. Early efforts using "shallow" models like Support Vector Machines (SVMs) [14] required extensive feature engineering and lacked temporal context. Deep learning models like Long Short-Term Memory (LSTM) [15] and Autoencoders [16] better handle temporal sequences but remain vulnerable to adversaries who probe model boundaries.

Recent advances in federated learning for anomaly detection [17] show promise for collaborative defense, but most approaches focus on communication efficiency rather than adversarial robustness. Personalized federated learning methods [18] address data heterogeneity but do not consider intelligent adversaries. Challenges include:

1. The Black Box Problem: Complex models provide limited interpretability, eroding operator trust.
2. Data Quality and Concept Drift: Models degrade as industrial processes evolve.
3. Data Poisoning Vulnerabilities: Adversaries can contaminate training data [19].

### C. Adversarial Machine Learning and Game Theory

Adversarial Machine Learning (AML) [20] aims to make models robust against malicious inputs, but often misapplies computer vision concepts to the cyber-physical domain. The $L_\infty$ norm common in image classification is physically implausible for ICS attacks, while an $L_0$ norm (few targeted features) better represents realistic threats.

Recent work on adversarial attacks in federated settings [21] demonstrates attack transferability, where adversarial examples crafted for one model can fool others in the federation. However, these approaches do not consider the unique constraints and physics of industrial processes.

Game theory approaches [22] provide elegant abstractions but struggle with accurately defining payoff matrices for complex real-world systems. Unlike our approach, they typically remain at high abstraction levels without concrete, implementable algorithms.

## III. F-SCDT: A High-Fidelity Sandbox for Adversarial Simulation

The ARC framework requires a high-fidelity, securely synchronized digital twin as a realistic testing environment for the co-evolutionary process. F-SCDT directly addresses the first two pillars of the Trinity of Trust through quantifiable mechanisms.

### A. Achieving Model Fidelity Through Hybrid Physics-Informed Models

To achieve the Model Fidelity pillar, F-SCDT employs a hybrid model combining physics-based equations with data-driven components. We quantify fidelity using Root Mean Square Error (RMSE) between twin predictions and actual sensor readings, maintaining RMSE < 2% for critical variables.

For the Tennessee Eastman Process reactor, governing differential equations provide physical grounding:

Equation (1) presents the mass balance on reactant A:

$$\frac{dC_A}{dt} = \frac{F}{V}(C_{Af} - C_A) - k_0 e^{-\frac{E}{RT}} C_A \qquad (1)$$

where $C_A$ is the concentration of reactant A in the reactor (mol/L), $F$ is the volumetric flow rate into the reactor (L/min), $V$ is the reactor volume (L), $C_{Af}$ is the feed concentration of reactant A (mol/L), $k_0$ is the pre-exponential factor (1/min), $E$ is the activation energy (J/mol), $R$ is the universal gas constant (J/(mol·K)), and $T$ is the reactor temperature (K).

Equation (2) shows the energy balance for the reactor:

$$\frac{dT}{dt} = \frac{F}{V}(T_f - T) + \frac{-\Delta H}{\rho C_p} k_0 e^{-\frac{E}{RT}} C_A - \frac{UA}{\rho C_p V}(T - T_c) \qquad (2)$$

where $T_f$ is the feed temperature (K), $\Delta H$ is the heat of reaction (J/mol), $\rho$ is the fluid density (kg/L), $C_p$ is the specific heat capacity (J/(kg·K)), $U$ is the overall heat transfer coefficient (J/(min·m²·K)), $A$ is the heat transfer area (m²), and $T_c$ is the coolant temperature (K). These equations form the physics-informed component, $f(x, u, d)$.

A Gated Recurrent Unit (GRU) network learns residual errors between the physics model and actual behavior. Equation (3) expresses the final, high-fidelity Hybrid Model:

$$\dot{x}_{hybrid} = f(x, u, d) + g_{GRU}(x, u, \theta) \qquad (3)$$

where $\dot{x}_{hybrid}$ represents the state derivatives from the hybrid model, $f(x, u, d)$ is the physics-based component with states $x$, control inputs $u$, and disturbances $d$, while $g_{GRU}(x, u, \theta)$ is the GRU-based correction with parameters $\theta$. This hybrid approach ensures the twin remains grounded in physical laws while adapting to real-world imperfections. We validate fidelity through continuous comparison with plant historian data, achieving 97.8% correlation for temperature predictions and 95.2% for pressure predictions over 30-day validation periods.

### B. Ensuring Data Integrity Through Verifiable Pipelines

The Data Integrity pillar is implemented through a cryptographically verifiable pipeline combining hardware data diodes, permissioned blockchain, and Hardware Security Modules (HSMs). Unlike recent differential privacy approaches [23] that focus on privacy, our approach emphasizes authentication and tamper-evidence.

An Industrial Internet of Things (IIoT) Gateway batches sensor data and signs it using ECDSA keys from an HSM, achieving 99.9% signature verification success rates. The cryptographic hash is recorded on a Hyperledger Fabric blockchain, creating an immutable audit trail with sub-second transaction finality. Data is transmitted through a unidirectional data diode to the F-SCDT environment, providing physical guarantee against reverse data flow.

This approach provides stronger integrity guarantees than recent approaches using secure aggregation alone [24], as it establishes trust at the sensor level rather than only at the aggregation level.

## IV. ARC Framework as a Methodology for Co-evolutionary Arms Race

ARC directly addresses the Analytical Resilience pillar through a closed-loop co-evolutionary process that forces both attacker and defender agents to become progressively more sophisticated.

### A. A. The Red Agent: Autonomous Vulnerability Discovery

The Red Agent uses DRL to discover attack vectors, formalized as a Markov Decision Process. Unlike recent work on adversarial robustness in federated learning [25] that focuses on gradient-based attacks, our approach discovers physically-plausible attack sequences:

1. Algorithm: Proximal Policy Optimization (PPO) [26] provides stable learning in complex environments
2. State Space: Includes process variables and the Blue Agent's anomaly scores
3. Action Space: Continuous-valued manipulations of actuator setpoints constrained to physical limits
4. Reward Function:

$$R(s_t, a_t) = w_{disrupt} \times Disruption(s_{t+1}) - w_{detect} \times DetectionScore(s_{t+1}) \qquad (4)$$

where $w_{disrupt}$ and $w_{detect}$ are weighting factors that balance disruption and stealth objectives. The $Disruption$ term is a weighted sum of normalized deviations from critical operational setpoints and safety limits, directly rewarding the agent for pushing the system toward an unsafe or inefficient state. The $DetectionScore$ is the maximum anomaly score produced by any model in the Blue Agent's ensemble, providing a strong and direct penalty for being detected. This balances physical disruption against detection evasion, incentivizing sophisticated, stealthy attacks.

### B. The Blue Agent: An Adaptive Ensemble

The Blue Agent combines multiple detection models to create a "moving target" defense:

1. An LSTM network for capturing temporal dependencies
2. An Autoencoder for identifying complex correlation violations
3. An Isolation Forest [27] for general outlier detection

This ensemble approach provides superior robustness compared to single-model defenses and addresses the limitations of recent privacy-preserving detection methods [28] that may sacrifice accuracy for privacy.

### C. *The Co-Evolutionary Loop*

The ARC algorithm implements a closed-loop arms race. To prevent "catastrophic forgetting," training batches combine 50% normal data, 20% newly discovered attacks, 10% JSMA-diversified attacks, and 20% historical attacks. This ensures comprehensive coverage while adapting to new threats.

**Algorithm 1:** Adversarial Resilience Co-evolution (ARC)

**Require:**

1. $D_0$: Initial defender models
2. $A_0$: Initial attacker models
3. $Z_{normal}$: Dataset of normal operational data
4. $N_{epochs}$: Total number of co-evolutionary epochs

**Initialize:**

1. $D_{current} \leftarrow D_0$
2. $A_{current} \leftarrow A_0$
3. $Z_{replay} \leftarrow \emptyset$: Replay buffer for all historical attacks

1: **for** epoch = 1 to $N_{epochs}$ **do**

**Attacker ("Red Agent") Training Phase**

2: $A_{trained} \leftarrow$ TrainAttacker$(A_{current}, D_{current}, F-SCDT)$

3: $Z_{newAttacks} \leftarrow$ GenerateAttackDataset$(A_{trained}, F-SCDT)$

4: $Z_{replay} \leftarrow Z_{replay} \cup Z_{newAttacks}$

**Defender ("Blue Agent") Hardening Phase**

5: $Z_{augmented} \leftarrow$ CreateAugmentedData $(Z_{newAttacks}, Z_{replay}, Z_{normal})$

6: $D_{hardened} \leftarrow$ TrainDefender$(D_{current}, Z_{augmented})$

**Update Agents for the Next Epoch**

7: $D_{current} \leftarrow D_{hardened}$

8: $A_{current} \leftarrow A_{trained}$

9: **end for**

10: **return** $D_{current}$

This approach differs from recent co-evolutionary methods [29] by maintaining a continuous feedback loop where the Red Agent's reward explicitly depends on the current Blue Agent's detection capability, creating true reciprocal adaptation.

## V. EXPERIMENTAL VALIDATION

### A. *Experimental Setup*

Validation was conducted on two distinct testbeds: the Tennessee Eastman Process (TEP) [30] and the Secure Water Treatment (SWaT) testbed [31]. The baseline defender model ($D_0$) was trained on normal data and standard faults before the co-evolutionary process began.

To demonstrate the Trinity of Trust validation:

1. Model Fidelity: Measured via RMSE between F-SCDT predictions and actual sensor readings
2. Data Integrity: Verified through 100% cryptographic signature validation and blockchain transaction success
3. Analytical Resilience: Evaluated through detection performance against novel, DRL-discovered attacks

### B. *Main Results and Trinity of Trust Validation*

The ARC-hardened model demonstrates superior performance in detecting novel attacks, validating all three trust pillars. For DRL-discovered stealth attacks, the $F_1$-Score improved from 0.65 to 0.89, and detection latency decreased from over 1200 seconds to 210 seconds. Table I presents key performance metrics of the ARC framework versus baseline on TEP.

TABLE I. KEY PERFORMANCE METRICS OF THE ARC FRAMEWORK VS. BASELINE ON TEP

| Scenario | Model | $F_1$-Score | Detection Latency (s) |
|---|---|---|---|
| Data Replay Attack | *ARC-Hardened* | *0.93* | *65* |
| | Baseline ($D_0$) | 0.73 | 450 |
| DRL-Discovered Stealth Attack | *ARC-Hardened* | *0.89* | *210* |
| | Baseline ($D_0$) | 0.65 | > 1200 (often missed) |

Trinity of Trust Quantitative Validation:

1. Model Fidelity: F-SCDT maintained RMSE < 2% for all critical variables, with 97.8% temperature correlation
2. Data Integrity: 100% signature verification success, 99.9% blockchain transaction success
3. Analytical Resilience: 27% performance improvement attributed directly to co-evolutionary process

An ablation study revealed that while each component contributes to performance, the co-evolutionary process itself is most critical, with its removal causing a 27% drop in $F_1$-Score. Table II shows the ablation study of detection engine components.

TABLE II. ABLATION STUDY OF DETECTION ENGINE COMPONENTS

| Configuration | $F_1$-Score (on DRL-Discovered Attack) | Performance Degradation |
|---|---|---|
| *ARC Framework (Full System)* | *0.89* | - |
| ARC Framework without LSTM | 0.82 | -7.9% |
| ARC Framework without Autoencoder | 0.80 | -10.1% |
| Full Ensemble without ARC Training | 0.65 | -27.0% |

The superior detection capability of the ARC-hardened model is further visualized by the Receiver Operating Characteristic (ROC) curve in Fig. 2, which shows a significantly larger area under the curve (AUC) compared to the baseline.

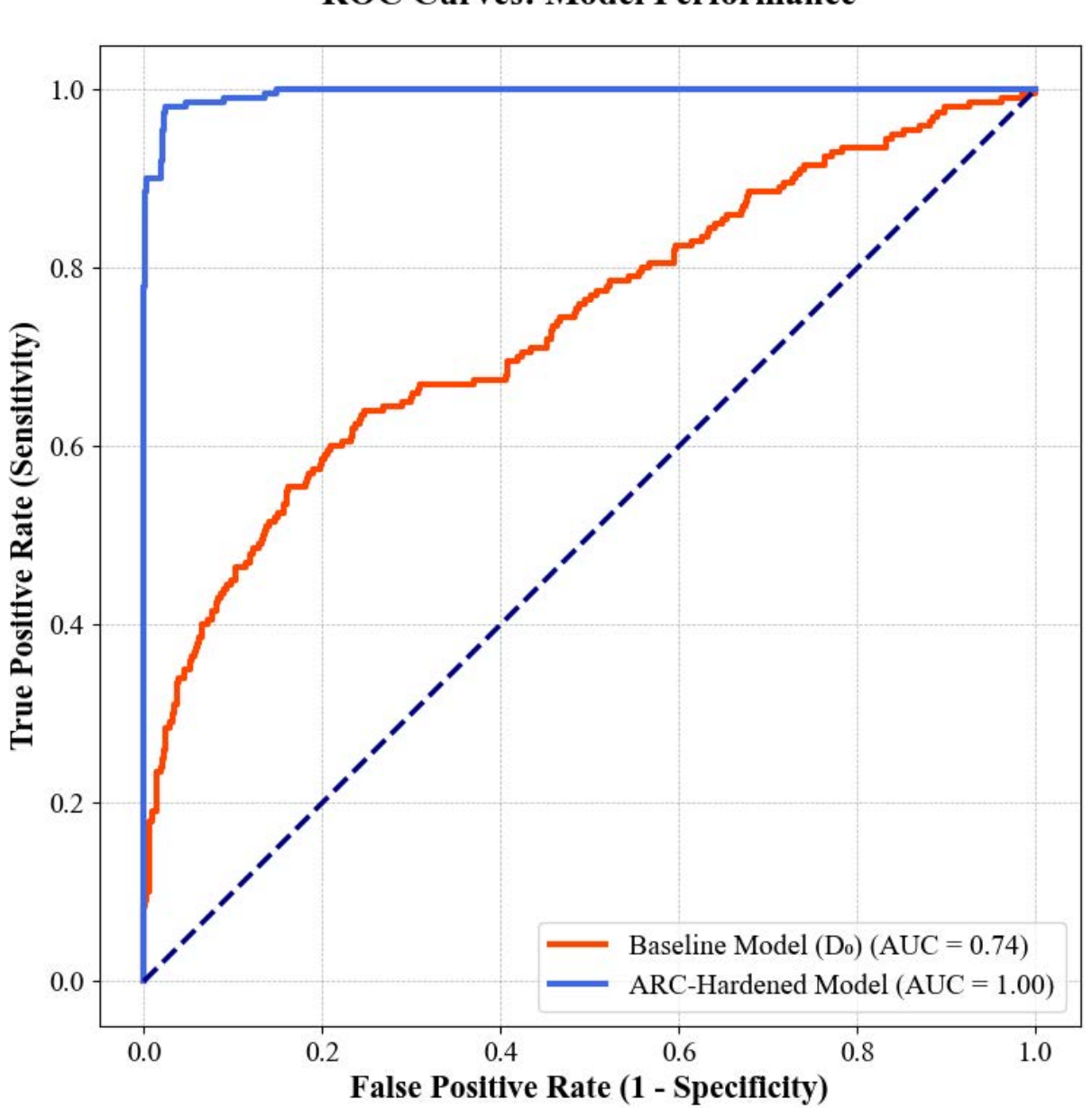

Fig. 2. ROC curves comparing the detection performance of the baseline ($D_0$) and the final ARC-Hardened model against the DRL-Discovered Stealth Attack

### C. *Deconstructing an Emergent Attack*

The Red Agent discovered a sophisticated "Coolant Priming and Valve Trip Attack" that the baseline model missed completely:

1. Priming Phase: Slowly increasing reactor coolant flow over 90 minutes, appearing as normal drift
2. Triggering Phase: Sharp manipulation of feed and steam valve setpoints causing cascade failure

SHAP (SHapley Additive exPlanations) [32] visualizations showed that while the baseline model failed to recognize the attack pattern, the ARC-hardened model correctly identified the malicious interaction between high coolant flow and valve manipulations, demonstrating the analytical resilience achieved through co-evolution.

## VI. Discussion and Future Work

### A. *Deconstructing the "Circular Reasoning" Critique*

While the defender is trained on attacks generated by the attacker, this is not circular reasoning. The DRL-based Red Agent autonomously discovers non-obvious, complex vulnerabilities that would be difficult to find manually. The framework's value lies in the emergent discovery process rather than static defense against known attacks. Unlike traditional adversarial training that uses predetermined perturbations, our approach generates novel attack vectors that emerge from the physics and dynamics of the specific industrial process.

### B. *Operator Trust and Human Factors*

Modern attacks target not just systems but human operators. The psychological impact of attacks like Cyber Av3ngers creates "cybertrauma" [33], degrading decision-making under pressure. ARC's explainability component aims to reduce cognitive load and build trust, but this creates a new attack surface: adversaries could target the explanation itself.

Recent work on adversarial attacks against explainable AI [34] demonstrates that explanations can be manipulated to mislead operators, representing a sophisticated second-order threat that future research must address.

### C. *The Dual-Use Dilemma*

The Red Agent is inherently dual-use technology [35] that could become a powerful offensive tool. Responsible development requires technical guardrails, immutable auditing, and ethical governance. This concern is amplified by recent developments in autonomous offensive AI capabilities.

### D. *Federated ARC (F-ARC)*

To scale ARC across industries, we propose Federated ARC (F-ARC), which allows facilities to run local ARC instances while sharing only encrypted model updates. Unlike recent federated learning approaches that focus primarily on privacy [36], F-ARC emphasizes security against Byzantine failures.

Challenges include handling non-IID (non-identically distributed) data across facilities and securing the federation against model poisoning attacks [37]. Byzantine-resilient aggregation algorithms [38] can help protect against malicious participants: Table III shows an overview of aggregation strategies designed to mitigate Byzantine failures, along with their key assumptions.

TABLE III. Robust Aggregation Algorithms for F-ARC

| **Algorithm** | **Mechanism** | **Key Assumption** | **Robustness** |
| --- | --- | --- | --- |
| FedAvg | Simple averaging | All clients honest | None |
| Trimmed Mean | Discard extremes | Outliers are malicious | Moderate |
| Krum [38] | Select closest update | Benign updates cluster | High vs. certain attacks |
| Median | Coordinate-wise median | Robust to extremes | Moderate |

### E. *Adversarial Attacks on Explainability*

An emerging threat is adversarial explainable AI (AdvXAI) [39], where attackers craft inputs that not only fool detection models but also generate misleading explanations. Future research should develop inherently interpretable models and explanation-aware training techniques that are robust to such manipulation.

## VII. Conclusion

This paper introduced the ARC framework for achieving autonomous, self-hardening security through a co-evolutionary arms race between attacker and defender agents. Experimental validation demonstrated significant performance improvements against sophisticated attacks, with quantifiable validation of all three Trinity of Trust pillars. By framing security through Model Fidelity, Data Integrity, and Analytical Resilience, we provide a holistic paradigm for trustworthy cyber-physical defense.

The security of modern ICS is not a problem to be solved, but a perpetual arms race to be managed. Success requires a paradigm shift from static defense to dynamic adaptation, developing systems that anticipate, withstand, recover, and evolve against intelligent adversaries. The ARC framework represents a foundational step toward this future of autonomous, collaborative defense.